\documentclass[manuscript]{aastex}

\newcommand{\etal}{\it et al.}
\pdfoutput=1

\slugcomment{Not to appear in Nonlearned J., 45.}

\shorttitle{Dwarf Galaxy Morphology}

\shortauthors{Ann}

\begin{document}

\title{Morphology of Dwarf Galaxies in Isolated Satellite Systems}

\author{H. B. Ann\altaffilmark{1} }
\affil{Department of Earth Science, Pusan National University,
    Busan, Korea}

\email{hbann@pusan.ac.kr}

\begin{abstract}
The environmental dependence of the morphology of dwarf galaxies in isolated
satellite systems is analyzed to understand the origin of the dwarf galaxy
morphology using the visually classified morphological types of 5836
local galaxies
with $z \lesssim 0.01$. We consider six sub-types of dwarf galaxies,
dS0, dE, dE$_{bc}$, dSph, dE$_{blue}$, and dI, of which the first four
sub-types are considered as early-type and the last two as late-type. The
environmental parameters we consider are the projected distance from the host
galaxy ($r_{p}$), local and global background densities, and the host
morphology. The spatial distributions of dwarf satellites of early-type
galaxies are much different from those of dwarf satellites of late-type
galaxies, suggesting the host morphology combined with $r_{p}$ plays a
decisive role on the morphology of the dwarf satellite galaxies. The local
and global background densities play no significant role on the morphology
of dwarfs in the satellite systems hosted by early-type galaxies. However,
in the satellite system hosted by late-type galaxies, the global background
densities of dE and dSph satellites are significantly different from those of
dE$_{bc}$, dE$_{blue}$, and dI satellites. The blue-cored dwarf
satellites (dE$_{bc}$) of early-type galaxies are likely to be located
at $r_{p} > 0.3$ Mpc to keep their cold gas from the ram pressure stripping by
the hot corona of early-type galaxies.
The spatial distribution of dE$_{bc}$ satellites of early-type galaxies and
their global background densities suggest that their cold gas is intergalactic
material accreted before they fall into the satellite systems.
\end{abstract}


\section{Introduction}

Dwarf galaxies are the most dominant populations of the nearby
universe ($z \lesssim 0.01$) as well as the building blocks of massive 
galaxies in the $\Lambda$CDM cosmology.
There are two types of dwarf galaxies: dwarf
elliptical-like galaxies (dEs) and dwarf irregular galaxies (dI). The dEs
include dS0, dE, dSph, dE$_{bc}$ and dE$_{blue}$ \citep{ann15}.
The dE$_{bc}$ represents dE galaxies with blue cores and the dE$_{blue}$ 
represents globally blue dwarf galaxies that have ellipsoidal shapes. Like the
blue elliptical galaxies \citep{str01}, dE$_{bc}$ and dE$_{blue}$ can
be distinguished from other dEs by the colors, not by the shapes.
The dSph is not always distinguished from the dE in literature
\citep{san84, fer94} but most previous studies recognize the dSph galaxy as 
a distinct class of galaxies \citep{mat98, gal94, gre97, vdb99}.

The morphology of a dwarf galaxy is easily affected by its environment because 
of its small mass. The environmental quenching, a sudden 
shutdown of star formation driven by the environment, 
is most pronounced for the 
dwarf galaxies in groups and clusters \citep{pen10, pen12, kov14,tal14,wet15}.
This is the reason why environmental quenching is sometimes termed  
satellite quenching. There are several mechanisms proposed for
environmental quenching \citep[see][for a review]{bosgav06}. This quenching
is most effective in cluster galaxies because the
density and temperature of the intracluster medium is high enough to remove 
gas from late-type galaxies by ram pressure stripping \citep{gun72,
roe05,jac07,boseli08,mcC08,ton09,bek09,vol09,book10,tecce10,kimm11,vij15,zinger16} and tidal 
interactions among galaxies are expected to be frequent enough to transform
their morphology via galaxy harassment \citep{moo98}. 
Morphology transformation by harassment is not effective in poor groups and 
satellite systems because encounters among satellite galaxies are too rare
to perturb their mutual orbits \citep{may01, pas03}. However, ram pressure
stripping seems to be operating in poor groups and satellite 
systems \citep{mar03,hes06,ras06,kaw08,gat13,fil16,eme16,bro17}. 

The correlation between the observed properties such as morphology, colors, and
star formation rate of central galaxies and their neighbors, dubbed as
"galactic conformity", is now well established \citep{wein06, ann08,kau13}.
In particular, \citet{ann08} found the morphology conformity between 
an isolated host and its satellites using morphology itself rather 
than proxies of galaxy morphology such as colors and star formation 
rate \citep{wein06, kau13}. The morphology conformity is thought to be
a result of morphology transformation driven by environmental quenching which
leads to a higher fraction of early-type satellites near early-type 
hosts than late-type hosts. The main mechanism of the environmental 
quenching in satellite systems that drives the morphology conformity is
the ram pressure stripping. It is more effective in early-type hosts  
because the hot corona of an early-type galaxy is larger 
than that of a late-type galaxy \citep{jel08,sul01,mul10,ras09}.
Recent models for satellite infall and ram pressure stripping \citep{sla14} 
show that quenching occurs within $\sim2$Gyr for low-mass satellites while
massive satellites continue to form stars for a prolonged time, $>5$Gyr after
falling into their host. Their models are consistent with the observations of
the local group (LG) that nearly all the dwarf galaxies 
with $M_{star} \lesssim 10^{9}M_{\odot}$ within 300 kpc from the Milky Way 
and M31 show no star formation \citep{wet15}.

In order to transform late-type galaxies into early-type ones, 
quenching of star formation is not sufficient. It should accompany structural 
changes, from disks to spheroidals, if dE and dSph galaxies are transformed 
from late-type disk galaxies. In satellite systems, tidal stirring by the host
galaxy plays a major role to transform dwarf irregular galaxies into dwarf
spheroidal galaxies \citep{may01,may06,lokas10,yoz15}. In cluster environment,
harassment \citep{moo98}
and tidal shock \citep{may01} are responsible for the structural changes 
after satellite galaxies enter into clusters or satellite systems. 
Tidal shocks also play some role in removing the cold gas from satellite
galaxies \citep{may06} but complete removal of cold gas requires hydrodynamical 
processes such as ram pressure stripping \citep{gun72}. 

Since the morphology of a galaxy depends on the environment where it forms,
we expect a close relationship between the satellite morphology and its
environment, parameterized by the local and global background
densities as well as the projected distance from the host galaxy ($r_{p}$) and
the host morphological type. It is of interest to see whether the spatial 
distributions of dwarf satellites can be distinguished by their sub-types.
Until now, there is no detailed study on the environment of dwarf galaxies, 
except for those in the Local Group \citep{vdb94},
local volume within 10 Mpc \citep{kar14} and the Virgo cluster \citep{lis07}.
The reason for the limited number of studies on the environment of the 
dwarf galaxies is that there is no redshift survey to cover the faint dwarf 
galaxies before the Sloan Digital Sky Survey \citep[SDSS;][]{yor00}.

In this regard, this study seeks to find some clues about the origin of dwarf
galaxies by examining the  spatial
distributions of satellite galaxies in the isolated satellite systems where the 
ram pressure from the hot intergalactic medium is thought to be too weak to
completely remove the cold gas in late-type galaxies because isolated satellite
systems are likely to be located in the low-density regions. If the morphology
of a galaxy can be transformed after it becomes a satellite galaxy, we expect
some correlations between satellite morphology and the distance
from the host galaxy because tidal and hydrodynamical interactions
depend on the distance between the host and its satellite.
Of course, in cases where morphology transformation is still in progress, they
are likely to be located in the outlying regions of satellite systems.
The rotationally supported dE galaxies which are located in the outer parts of
the Virgo cluster \citep{tol11} may be examples of dE galaxies transformed 
from late-type disk galaxies. \citet{ben15} showed that tidally stirred 
dE galaxies with significant rotation are located in the outer parts of the 
model clusters.

The dE$_{bc}$ galaxy is of special interest. 
The pronounced feature distinguishing dE$_{bc}$ 
galaxies from other dEs is the blue core where active star formation 
is supposed to be taking place, suggesting the presence of cold gas there.
Then, what is the origin of the cold gas? It could be the intergalactic 
material accreted onto a dE satellite or leftover material after being 
transformed from late-type galaxies. It is also possible that the cold gas of
a late-type host can be transferred to a dE to make a dE$_{bc}$. There are no 
simulations to predict such a cold gas transfer from a late-type host to its
satellite although hydrodynamical simulations, e.g., distant encounter models 
of \citet{hwang15}, show that cold gas in a late-type galaxy can be
transferred to the center of an early-type galaxy whose mass is comparable to
the late-type galaxy. We will explore the origin of the cold gas in dE$_{bc}$
satellites by analyzing their environment. 

This paper is organized as follows. The data and sample selection are 
described in section 2 and the results of the present study are given in
section 3. Summary and discussion on the origin of the dwarf galaxy 
morphology are given in the last section.

\section{Data and Sample Selection}

\subsection{Morphological Types of Dwarf Galaxies}

We used the galaxies in the Catalog of Visually Classified Galaxies in the
nearby universe \citep[CVCG;][]{ann15} as the primary sample 
of the present study. The basic source of data in the CVCG is the 
Korea Institute of
Advanced Study Value-Added Galaxy Catalog \citep[KIAS-VAGC;][]{chk10} which
provides photometric and spectroscopic data derived from the SDSS Data
Release 7 \citep{ab09}. They supplemented spectroscopic redshifts of galaxies
brighter than the magnitude limit ($r_{petrosian} \sim 14.5$) of the target 
galaxies of the SDSS spectroscopic observation from
various redshift catalogs. \citet{ann15} added faint galaxies from the NASA
Extragalactic Data Base (NED) and the galaxies in \citet{mk11} that
are not overlapped with the galaxies from the KIAS-VAGC. These supplements make
the CVCG nearly complete to galaxies brighter than $r_{petrosian}=17.77$.

The detailed description of the morphological properties of dwarf galaxies is
given in \citet{ann15} but here we describe the morphology of dwarf galaxies
briefly. 
For three sub-types of dE-like galaxies (dS0, dE, dSph), the presence/absence
of nucleation is distinguished by the subscript 'n' and 'un'.  For example,
the dE$_{n}$ galaxy represents the dE galaxy with nucleation while dE$_{un}$
denotes the dE galaxy with no nucleation. However, we do not
distinguish nucleated dwarfs from un-nucleated ones in the present study
because of the small sample size of the dwarf satellites.

Figure 1 shows the prototypical images of the dwarf elliptical-like galaxies. 
As can be seen in the first and second rows of Figure 1, dS0 and 
dS0$_{p}$(peculiar dwarf lenticulars) galaxies are characterized by the 
lens-like features in the central regions.
Some of them show signatures of spiral arm remnant in the outer part of the
disk \citep{jer00,bar02,geh03,gra03,rij03,lis06,lis09,jan12}. DS0 with
spiral arms are denoted as dS0$_{p}$ to distinguish them from the normal dS0s.
Because of the small number, we do not distinguish dS0 and dS0$_{p}$ in the
analysis of the correlation between the satellite morphology and environment.
The dE galaxies in the third row of Figure 1 show the basic morphological
properties of the dwarf elliptical-like galaxies, characterized by the 
elliptical shape, shallow surface brightness gradient, presence/absence of 
nucleation, and some signatures of spiral arm remnant (in the third column).
The dE$_{bc}$ galaxies in the fourth row are characterized by the blue core
which manifests the presence of young stellar populations there. Except for
the blue core, they are almost identical to dE galaxies. While the dE$_{bc}$
galaxies show blue colors in the core only, the colors of the dE$_{blue}$
galaxies are globally blue, similar to that of dwarf irregular (dI) galaxies.
As shown in the 6th row of Figure 1, the dE$_{blue}$ galaxies have round shapes
resembling the HII region-like blue compact dwarfs (BCDs) in appearance.
But their colors are less bluer than BCDs, implying less star formation.  
The dE$_{blue}$ and BCD can be classified as the same type if BCDs embedded in 
amorphous irregular galaxies are excluded.
The dSph galaxies look similar to dE galaxies 
but they are less luminous and somewhat bluer than dE galaxies. 
Because of the similar morphology of dE and dSph galaxies, 
they are often used interchangeably in the literature. From a photometric point
of view, their colors are similar to each other but dE is slightly brighter
than dSph. However, from the kinematic point of view, dE galaxies are mostly
supported by rotation \citep{dav83, ped02} while dSph galaxies are supported
by dispersion \citep{geh02}.

We used galaxy distances derived from the redshifts corrected for 
the motion relative to the centroid of the LG using the prescription of
\citet{mou00}. If NED provides galaxy distances determined from the distance 
indicators, these distances were used.
The number of galaxies with their distances from distance indicators
is 1548. The galaxies lying inside a $10^{\circ}$-cone around M87 with a 
redshift less than $z = 0.007$ \citep{kra86} are assumed to be the members
of the Virgo Cluster, and we used the distance of the Virgo
cluster for these galaxies. We assumed the Virgo distance as $D=16.7$ Mpc and
H=75 km s$^{-1}$ Mpc$^{-1}$.

\begin{figure}
\includegraphics{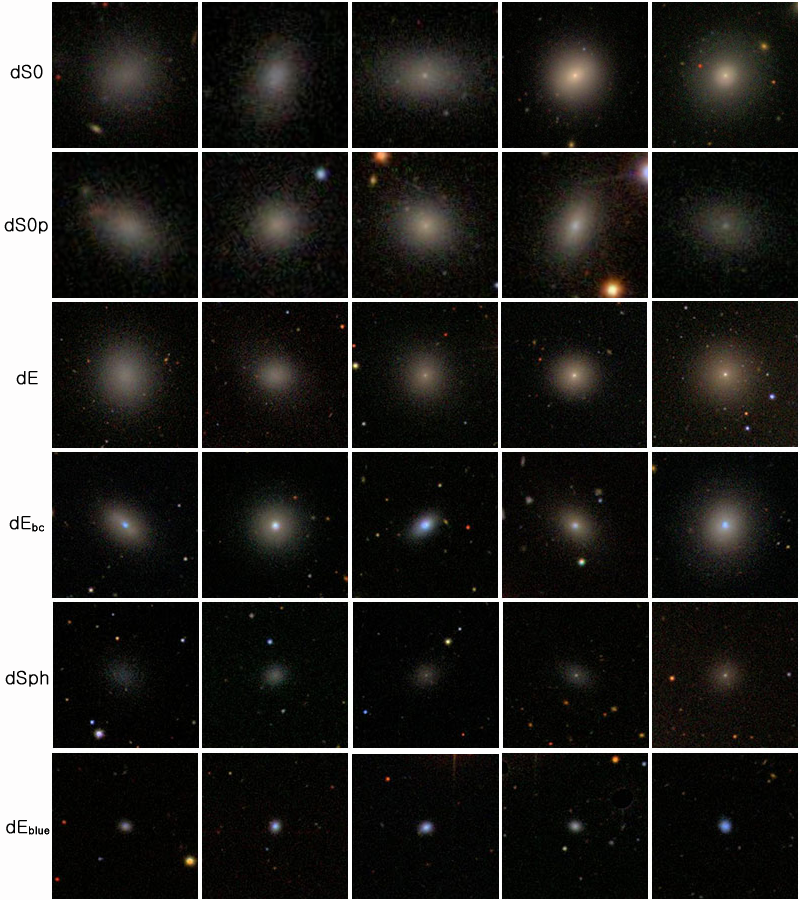}
\caption{Sample images of dwarf elliptical-like galaxies. For dS0 (dS0$_{p}$),
dE, and dSph galaxies, galaxies with no nucleation are displayed in the 
first two columns from the left.\label{fig1}} 
\end{figure}


Figure 2 shows the histograms of $M_{r}$ of six sub-types of dwarf galaxies 
in the CVCG.
We derived $M_{r}$ from the $r$-model magnitude corrected for the 
extinction in the Galaxy using the $r$-band extinction given in the SDSS DR7.
We did not apply K-correction and evolution correction. As can be seen in
Figure 2, the luminosity distributions of the six sub-types of dwarf galaxies 
are quite different. Four of them (dS0, dE, dE$_{bc}$ and dI) show skewed 
distribution with different peak luminosities and the  
two (dSph and dE$_{blue}$) show a roughly Gaussian shape. The majority of dwarf 
galaxies are fainter than $M_{r}=-17$ with some fractions of galaxies brighter
than $M_{r}=-17$ in dS0 ($15\%$), dE ($10\%$) and dE$_{bc}$ ($16\%$) galaxies.
The reason for the high fraction of bright galaxies ($M_{r}< -17$) in 
dE$_{bc}$ galaxies is due to the presence of the bright blue cores. There is
a significant difference in the luminosity distributions between dE and dSph
galaxies. On average, dE galaxies are $\sim1.5$ mag brighter than dSph 
galaxies.

\begin{figure}
\includegraphics{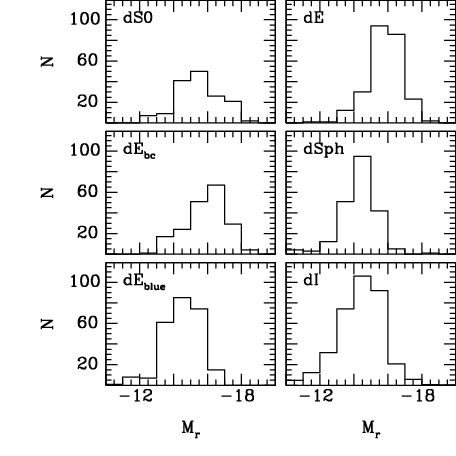}
\caption{Luminosity distributions of six sub-types of dwarf galaxies. 
In case of dI galaxies, the numbers of galaxies in each luminosity bin are 
one fourth of the observed numbers of galaxies in each bin.\label{fig2} }
\end{figure}


\subsection{Selection of Isolated Satellite Systems}

A galactic satellite system which consists of a host and its satellites
is said to be isolated if its host galaxy is isolated. 
The host galaxy of an isolated
satellite system, sometimes called a central galaxy, is the most massive 
galaxy in the satellite system which has no companion galaxy that has a
comparable mass. An isolated satellite system can be a part of a group or 
the group itself depending on the definition of group and isolation criteria.  
In this sense, the definition of an isolated satellite system adopted in
this study is nearly the
same as that of the group used by \citet{tul15}. However, the isolation
criteria are not unique in the literature. They depend on how to define
neighbor galaxies that have a comparable luminosity, $\delta M=1\sim2$mag.

In most cases, neighbor galaxies are defined by two parameters, linking
distance ($LD$) and linking velocity ($\Delta V^{\ast}$). The $LD$ is the 
maximum projected separation between a target galaxy
and its neighboring galaxy and $\Delta V^{\ast}$ is
the maximum radial velocity difference relative to the 
target galaxy. Most previous works used a fixed $LD$ such as 0.5 Mpc and
1 Mpc \citep{mck02, pra03, sal04, bos04, bos05, zen05, che06, sal07}. We
use variable $LD$ following \citet{ann08} who defined $LD$ as the sum of the 
virial radii of host and its neighbor galaxy. One of the merits using the sum
of the virial radii of host and satellite as $LD$ is that the number of 
interlopers can be minimized by constraining the boundary of satellite systems
as a physical boundary in which hydrodynamical interactions between host and
satellites are expected.

The constraint on $\Delta V^{\ast}$ is not simple because it should depend on
the environment of the satellite systems as well as the mass of host galaxy.
However, we assume a fixed value of $\Delta V^{\ast}$=500 km s$^{-1}$ for
all the host galaxies regardless of their mass and environment. This value 
of $\Delta V^{\ast}$=500 km s$^{-1}$ is different from that adopted 
by \citet{ann08} who searched isolated satellite systems of galaxies 
with $0.02 < z < 0.047$ and $M_{r}<-18$. They 
assumed $\Delta V^{\ast}$=1000 km s$^{-1}$ to constrain
the neighbor galaxies. 
We suppose that taking $\Delta V^{\ast}$=500 km s$^{-1}$
is a reasonable choice because the sample galaxies are confined to those
with $z \lesssim 0.01$ and the rms peculiar velocity of galaxies in the
local universe is roughly 500 km s$^{-1}$ \citep{pee79, pee87}.

Thus, we consider a galaxy with $M_{r}$ is isolated if there is no neighbor 
galaxy that has absolute magnitude brighter than $M_{r}+ \delta M_{r}$ with
$|{\Delta V}| < \Delta V^{\ast}$ within the projected separation from the 
host galaxy $r_{p}$ less than $r_{vir}$ + $r_{virnei}$ where $r_{vir}$
and $r_{virnei}$ are virial radii of the host and its neighbor galaxy, 
respectively. The virial radius of a galaxy is defined as the radius where the 
mean density within the sphere centered at the galaxy, 
$\rho=3\gamma L/4\pi r_{p}^{3}$,
becomes the virialized density, which is set to 766$\bar{\rho}$ where 
$\bar{\rho}$ is the mean mass density of the universe \citep{par08}.
We use the mass-to-light ratio ($\gamma$) determined from the $B-V$ colors 
\citep{wil13} which are derived from $g-r$ colors using the transformation 
equation $B-V = 0.98(g-r) + 0.22$ \citep{jes05}.

The number of isolated host galaxies selected from 5836 galaxies in the
local universe depends on $\delta M_{r}$. After examining several values 
of $\delta M_{r}$ we adopt $\delta M_{r}=1.0$ for the present study. 
It gives 346 isolated satellite systems. 
Once an isolated host is selected, it is easy to find satellite galaxies that 
belong to it. This is because all the galaxies satisfying the constraint 
for the neighbor galaxy are considered to be satellite galaxies if they 
are located within the $LD$ defined by the sum of the virial radii of the
host and satellite. We find 835 satellite galaxies that belong to 346 
isolated hosts. Among the 835 satellites, 411 satellites are dwarf galaxies.
Table 1 lists the number of satellites in six sub-types of dwarf galaxies 
sorted by the host morphology.
It seems worthwhile to note that the two sub-types, dS0 and dE$_{blue}$, 
behavior very differently. The majority ($\sim83\%$) of dS0 satellites belong
to early-type host while the majority ($\sim88\%$) of dE$_{blue}$ belong
to late-type host. The dE$_{bc}$ galaxies which we consider as early-types
show host distribution similar to the sub-types considered as 
late-types (dE$_{blue}$ and dI).

\begin{deluxetable}{crrrcrccrrcrr}
\tabletypesize{\scriptsize}
\tablecaption{Number of satellites in six sub-types of dwarfs.}
\tablewidth{0pt}
\tablehead{
\colhead{host} & \colhead{dS0\tablenotemark{a}}  & \colhead{dE }  & 
\colhead{dE$_{bc}$} & \colhead{dSph}  & 
\colhead{dE$_{blue}$\tablenotemark{b}} &  \colhead{dI} \\
}
\startdata
 49 E/S0    &  15 (4) & 13 &  8 & 10 &  5 (1) &  52 \\
297 Sp/Irr  &   3 (1) & 18 & 31 & 14 & 46 (9) & 176 \\
\enddata
\tablenotetext{a}{It includes dS0$_{p}$ galaxies whose number is given in the parenthesis}
\tablenotetext{b}{It inlcudes HII region-like BCDs 
whose number is given in the parenthesis.}
\end{deluxetable}

We used all the dwarf satellites in Table 1 to derive the surface density and 
early-type fractions of the satellite systems. However, we did not use the
satellite systems which are located close to the boundary of the SDSS survey 
volume in the derivation of the local background density because the local
background densities are supposed to be underestimated near the survey
boundary. There are 22 satellite systems whose distances to the 
survey boundary are closer than the distances to the $5$th nearest galaxies. 
The 22 satellite systems near the survey boundary are hosted by the late-type
galaxies. Among the 22 satellite systems, 9 satellite systems have
dwarf satellites. The total number of dwarf satellites in these 9 satellite
systems are 10. They are mostly dwarf irregular galaxies. The mean distance
to the $5$th nearest galaxies for these 9 satellite systems is $\sim1$ Mpc.

\section{Results}

We treat the isolated satellite systems selected by the previous section as 
an ensemble to examine the radial distribution of satellites because only a 
few satellite systems have satellites numerous enough to analyze their
spatial distribution. The average number of satellites in an isolated
satellite system is $\sim2.4$ and the number of dwarfs is similar to that
of giants.

\begin{figure}
\includegraphics[width=1\columnwidth]{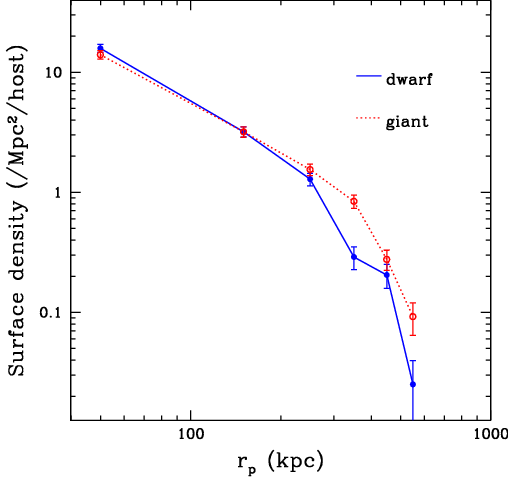}
\caption{Surface density distributions of dwarf and giant satellites along the
galactocentric distance ($r_{p}$) from host galaxy. 
The solid line represents dwarfs and the
dotted line indicates giants. Errors are Poisson errors.\label{fig3} }
\end{figure}


\subsection{Radial Distribution of Satellite Galaxies}
\subsubsection{Satellite Surface Density}

Figure 3 shows the distributions of surface density of dwarf and 
giant satellites as a function of projected distance from the host galaxy, 
i.e., galactocentric distance, $r_{p}$.
The surface density plotted in Figure 3 is the number of satellites in
each annulus divided by the annulus area and the number of host galaxies.
The width of each annulus is 100 kpc. At first glance, the surface density of
satellites decreases outward rapidly with a sharp boundary at 
$r_{p}\approx600$ kpc. The general trend of the surface densities of dwarf and
giant satellites is similar to each other except for the more rapid decrease
of dwarfs at $r_{p} > 250$ kpc. The reason for the significantly smaller number
of dwarfs than giants at $r_{p} > 250$ kpc is a natural consequence of the 
constraint imposed on searching for satellites. We used a variable $LD$
that depends on the virial radii of host and satellite.
That is, the limiting separation between the host and its satellite is set by
the sum of their virial radii. Since dwarfs have smaller virial radii than 
giants, they can not be located at large distances from the host galaxy.
To check this,  
we searched satellites by assuming a fixed radius of 500 kpc as the
boundary of a satellite system and found that the number of dwarfs in the 
outer part is close to or larger than the number of giants. However, we suppose
that the majority of dwarf satellites found beyond the $LD$ 
are field galaxies.

Figure 4 shows how the spatial distributions of dwarf satellites vary
with the host morphology. We use the $r_{p}$/R$_{vir}$ to represent the 
galactocenteric distance of a satellite galaxy.
First of all, what is remarkable is that the surface density of early-type
dwarfs is much higher than that of the late-type dwarfs in the innermost 
regions of the satellite systems hosted by early-type galaxies, while the
surface density of early-type dwarfs is lower than that of the the late-type
dwarfs in the outer parts of the satellite systems ($r_{p}$/R$_{vir} > 0.6$).
Here, we consider the
dE, dE$_{bc}$, dSph and dS0 galaxies as early-type dwarfs, whereas the dI, 
dE$_{blue}$ and the HII region-like BCDs as late-type dwarfs.  
Considering that the number of late-type dwarfs is much larger than the 
early-type dwarfs, the larger numbers of early-type dwarfs 
in the vicinity of early-type host galaxies imply that a significant fraction
of late-type dwarfs are transformed into early-type dwarfs by some mechanisms.
This can be ram pressure stripping by the hot corona of early-type host. 
Tidal stirring by the massive early-type host may play some role inshaping the
early-type morphology. On the other hand, in the vicinity of a late-type host,
late-type dwarfs are more abundant than early-type dwarfs. The reason for the
low surface density of early-type dwarfs near the late-type host is two-fold.
One is that the ram pressure from the hot corona of a late-type host is not
strong enough to remove the cold gas from late-type dwarfs because
the hot corona of late-type galaxies is thought to be smaller and less
luminous than that of early-type galaxies \citep{jel08,sul01,mul10, ras09}.
In general, X-ray bright galaxies are more massive and luminous than X-ray 
faint galaxies \citep{for85,tri85,fab89}. Thus, the ram pressure
stripping and tidal stirring are more effective in satellites belonging to 
early-type host galaxies. The other reason is that the cold gas in late-type 
hosts can be transferred to the satellite galaxies if they are close enough 
to exchange their gas. The high surface density of early-type satellites in
the central regions of the systems hosted by early-type galaxies and 
the low surface density of early-type satellites in the systems
hosted by late-type galaxies are the galaxy morphology conformity 
found in the satellite systems \citep{ann08}. One thing worth noting is that
the surface density of the dwarf satellites in the systems with early-type
hosts is higher than that of the satellite systems with late-type
hosts. The overall higher surface density around 
the early-type host galaxy is easy to understand if we consider the 
morphology-density relation \citep{dre80} and luminosity-density 
relation \citep{park07} which dictate that massive galaxies are likely to be 
formed in the high-density regions.

\begin{figure}
\includegraphics[width=1\columnwidth]{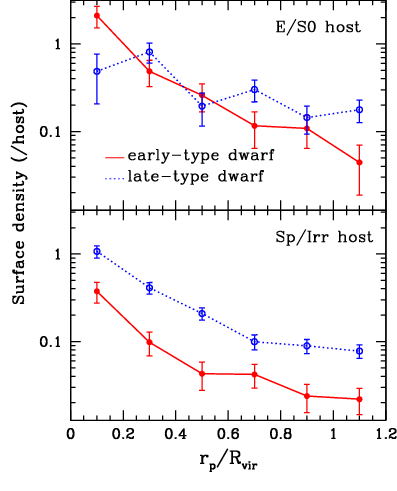}
\caption{Surface number density of dwarf satellites, sorted by host and
satellite morphology, as a function of $r_{p}/R_{vir}$. Satellite systems 
hosted by early-type galaxies are in the upper panel and those hosted by 
late-type galaxies are in the lower panel. Solid lines represent early-type
dwarf satellites and dotted line represent late-type dwarf 
satellites. \label{fig4}}
\end{figure}

\subsubsection{Early Type Fraction of Dwarf Satellites}

Since the finding that the morphology of a satellite galaxy is likely to
resemble the morphology of its host galaxy \citep{wein06, ann08, kau13},
morphology conformity becomes an important tool to understand the formation
of galaxies. The galaxy morphology conformity 
holds for any two galaxies which are near enough to interact
hydrodynamically \citep{par08}. However, galaxies in the previous studies
are mostly giant galaxies because of the observational data they used.
\citet{kau13} include dwarf galaxies in their sample
but it is limited to dwarf galaxies brighter than $M_{r}=-16$.
Here we present the morphology conformity between the giant
host and its dwarf satellites in the isolated satellite systems.
In the following, as in Figure 4, we consider the four sub-types of dwarf
galaxies (dS0, dE, dE$_{bc}$ and dSph) as early-types and the other
two sub-types (dE$_{blue}$ and dI) as late-types. The HII region-like BCDs
are combined with the dE$_{blue}$ galaxies because there is no significant
difference between them.

Figure 5 shows the early-type fraction $f_{e}$ of dwarf satellite galaxies
as a function of $r_{p}/R_{vir}$. It is apparent that the degree of morphology
conformity between a host and its satellites decreases with $r_{p}/R_{vir}$.
At $r_{p}/R_{vir}\approx 0.15$, the early-type satellite fraction of the
early-type host is $\sim0.6$ while that of late-type host
is $\sim0.2$. The degree of galaxy morphology conformity between the
host and its dwarf satellites revealed in the isolated satellite systems is
stronger than that found in \citet{ann08} where satellites are brighter than
$M_{r}=-18$. The difference in $f_{e}$ between early-type hosts and 
late-type hosts decreases with $r_{p}/R_{vir}$ and becomes indistinguishable
at $r_{p}/R_{vir} > 1.2$. 
\citet{ann08} interpreted the morphology conformity found in their sample
galaxies ($0.02 < z < 0.047$) as a result of the hydrodynamical interactions
between the hot corona gas of the host galaxy and the cold gas in the
satellite galaxies, which removes the cold gas from the satellites.

\begin{figure}
\includegraphics[width=1\columnwidth]{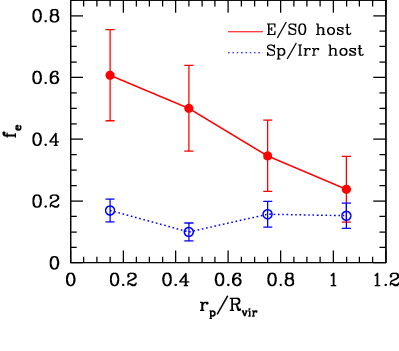}
\caption{Early-type fractions as a function of $r_{p}/R_{vir}$. The early-type
fractions of dwarf satellites hosted by early-type galaxies are plotted
by a solid line and filled circles while those of dwarf satellites hosted by 
late-type galaxies are plotted by a dotted line and open circles. \label{fig5}}
\end{figure}

\subsubsection{Cumulative Distribution of Dwarf Satellite Galaxies}

Since the number of satellite galaxies in the isolated satellite systems 
analyzed here is not large enough to suppress the statistical noise in 
the spatial distributions, we analyze their
cumulative spatial distribution along the projected galactocentric
distance ($r_{p}$). Figure 6 shows the cumulative fractions of
satellite galaxies grouped by the sub-types as a function of $r_{p}$.
The upper panel shows the cumulative spatial distributions of
satellite galaxies with early-type host galaxies, and the lower panel shows 
the cumulative spatial distributions of satellite galaxies with late-type 
host galaxies. The cumulative distribution of satellite galaxies shown in 
Figure 6 is notable in several respects. First of all, the cumulative 
distribution of the satellite galaxies along the projected distance varies 
greatly depending on the type of host galaxy. When the host galaxies are 
early-types, the cumulative distributions of satellites with different
dwarf morphology differ greatly, whereas they are less different, especially
in the central regions at $r_{p} < 0.1$ Mpc, when the host galaxies are  
late-types. 
Table 2 shows the probabilities of the K-S test to see the statistical 
significance of the differences in the spatial distributions of dwarf
satellite galaxies. We used the significance level of $\alpha=0.05$ to 
test the null hypothesis that the two samples are drawn from the same 
population. Thus, if the probability of the K-S test is
less than 0.05, the spatial distribution of the two samples is thought to
be statistically different. For example, for dwarf satellites of early-type
galaxies, the spatial distribution of dE satellites is statistically different
from those of other dwarf satellites except for dSph, whereas the spatial
distribution of dE$_{blue}$ satellites is statistically different from that
of dE satellites only. In cases of dwarf satellites of late-type host galaxies,
there is no statistical difference between the spatial distributions of 
dwarf satellite galaxies. 

\begin{table}{t}
\begin{center}
\centering
\caption{Probabilities of the K-S test for 6 sub-types of dwarf galaxies}
\doublerulesep2.0pt
\renewcommand\arraystretch{1.5}
\begin{tabular}{rcccccc}
\hline \hline
\multicolumn{7}{c}{Early type host} \\
\hline
Type & dS0 & dE & dE$_{bc}$ & dSph & dE$_{blue}$ & dI \\
\hline
 dS0 &  1.000 &   0.044 &   0.038 &   0.653 &   0.388 &   0.740\\
  dE  &   0.044 &   1.000 &   0.008 &   0.689 &   0.020 &   0.024\\
 dE$_{bc}$  &   0.038 &   0.008 &   1.000 &   0.047 &   0.181 &   0.048\\
 dSph &   0.653 &   0.689 &   0.047 &   1.000 &   0.375 &   0.633\\
dE$_{blue}$  &   0.388 &   0.020 &   0.181 &   0.375 &   1.000 &   0.355\\
 dI  &   0.740 &   0.024 &   0.048 &   0.633 &   0.355 &   1.000\\
\hline
\multicolumn{7}{c}{Late type host} \\
\hline
Type & dS0 & dE & dE$_{bc}$ & dSph & dE$_{blue}$ & dI \\
\hline
 dS0  &  1.000 &   0.690 &   0.722 &   0.259 &   0.540 &   0.409\\
 dE  &   0.690 &   1.000 &   0.470 &   0.123 &   0.219 &   0.161\\
 dE$_{bc}$  &  0.722 &   0.470 &   1.000 &   0.709 &   0.515 &   0.510\\
 dSph  &  0.259 &   0.123 &   0.709 &   1.000 &   0.708 &   0.601\\
dE$_{blue}$   &  0.540 &   0.219 &   0.515 &   0.708 &   1.000 &   0.834\\
 dI  &  0.409 &   0.161 &   0.510 &   0.601 &   0.834 &   1.000\\
\hline
\end{tabular}
\end{center}
\end{table}

In the case of the dE galaxies, the spatial distribution is 
most centrally concentrated in the
satellite systems hosted by early-type galaxies while they are less
centrally concentrated than dE$_{bc}$, dE$_{blue}$ and dI galaxies in the
satellite systems hosted by the late-type galaxy. The dSph satellite  
galaxies show a spatial distribution similar to that of the dE galaxy with a
slightly less central concentration with the early-type host
and a slightly higher central concentration with the late-type
host, especially at $r_{p} < 0.2$ Mpc. 
The difference in the spatial distributions of dE and dSph
galaxies in the satellite systems hosted by early-type galaxies and late-type 
galaxies is very natural if cold gas can be removed from satellite galaxies
or transferred from host to satellite by hydrodynamical interactions between
host and satellite \citep{raf15}. 
Cold gas in the progenitors of dE and dSph satellites 
lying close to the early-type host can be easily removed by interaction
with the early-type host which have a large hot corona to make dE and dSph
galaxies. The dE and dSph galaxies at a large distance 
from late-type hosts might
be formed with the present morphology and could keep their morphology because
of the weak interactions with the host galaxy. The removal of cold gas of
dE and dSph satellites by the hot halo gas in host galaxies, especially for
early-type hosts, seems to be consistent with the anti-correlation between the 
HI content and the halo mass \citep{sta16, bro17}.

\begin{figure}
\includegraphics[width=1\columnwidth]{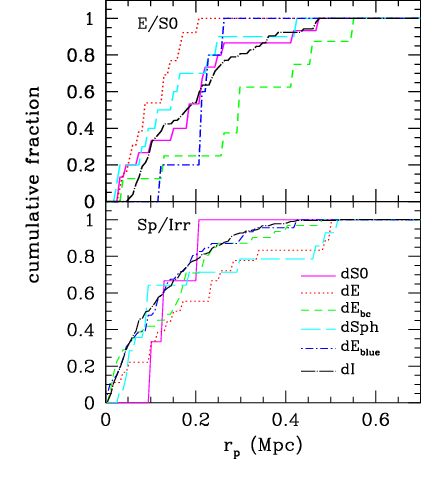}
\caption{Cumulative fraction of dwarf satellites as a function of $r_{p}$.
Satellite systems hosted by early-type galaxies are in the upper panel and 
those hosted by late-type galaxies are in the lower panel.
The cumulative fractions of the five sub-types of dwarf elliptical-like
galaxies and dwarf irregular galaxies are plotted by different lines:
solid line (dS0), dotted line (dE), short-dashed line (dE$_{bc}$),
long-dashed line (dSph), dot and short-dashed line (dE$_{blue}$), and
dot and long dashed line (dI). \label{fig6}}
\end{figure}

The spatial distribution of dE$_{bc}$ galaxies is of special interest. In the
satellite systems hosted by early-type galaxies, the dE$_{bc}$ 
galaxies show a spatial distribution which is significantly different from that
of other dwarf galaxies. The cumulative fractions of other dwarf satellites
in early-type hosts are greater than $\sim60\%$ at $r_{p}\approx0.2$ Mpc, 
whereas the cumulative fraction of dE$_{bc}$ galaxies is only $\sim25\%$
at the same radius. More than half of the dE$_{bc}$ satellites of early-type
galaxies are located at $r_{p}>0.25$ Mpc.
Their spatial distribution is thought to be closely related 
to their structural properties, i.e., young stellar populations at the center
of galaxies, indicating current star formation and cold gas there. 
However, since the amount of cold gas
in dE$_{bc}$ satellites is not supposed to be large enough to maintain star 
formation for a long time, it seems plausible that the cold gas
is recently accreted from the outside similar to some dE$_{bc}$ galaxies 
in the Virgo cluster \citep{hal12}. The presence of young stellar
populations in the central regions of dE$_{bc}$ galaxies, imposes some
constraints on their environment. They should be located in the regions far
from the early-type host galaxies to avoid the ram pressure stripping.
In cases of the dE$_{bc}$ satellites of late-type galaxies, they can be located
close enough to be affected by the ram pressure of the hot corona of host 
galaxies because cold gas can be transferred from the late-type host. 
Moreover, the ram pressure from the hot corona of late-type galaxies is 
less than that of the early-type galaxies because their X-ray luminosity is
less luminous than early-type galaxies \citep{for85,tri85,fab89}.
The spatial distribution of dE$_{bc}$ galaxies shown in Figure 6 seems to
supports the constriant for the location of dE$_{bc}$ satellites to keep
their cold gas.

The dE$_{blue}$ galaxies in the satellite systems hosted by early-type
galaxies show a very narrow range of distribution. They exist only at
0.1 Mpc $< r_{p} <0.3$ Mpc. Since the number of dE$_{blue}$ galaxies in these
systems is so small, only five galaxies, the narrow spatial distribution may
be due to the poor statistics, However, the absence of dE$_{blue}$ galaxies
at $ r_{p} < 0.1$ Mpc seems to be a real feature because cold gas in a small
system such as a dE$_{blue}$ galaxy can not survive inside the hot dense
corona of early-type galaxies. This is plausible because the hot
corona of early-type galaxies seems to extend to $\sim0.1$ Mpc \citep{for85}.
On the other hand, the absence of dE$_{blue}$ galaxy at  $r_{p} > 0.3$ Mpc 
is simply due to the small number statistics. In the satellite systems hosted
by late-type galaxies, the distribution of dE$_{blue}$ galaxies is nearly
identical to that of dI galaxies and similar to that of dE$_{bc}$ galaxies.
The common property of these three sub-types of dwarf galaxies is the presence
of cold gas.

\subsection{Background Density Dependence}

As we have seen in the previous section, the spatial distributions of satellite 
galaxies around host galaxy depend on the morphology of host and satellite.
The dependence of satellite morphology on the galactocentric
distance ($r_{p}$) seems to be related to the local background density while
the dependence of satellite morphology on the host morphology is related to 
the global background density. This is  
because there is a clear correlation between $r_{p}$ and the local background
density in groups and clusters \citep{dre80, wg91} and host morphology depends
on the background density, probably global density, via the morphology-density
relation \citep{dre80}.

The background density of a galaxy can be defined by several 
ways \citep{mul12}. We adopt the $n$th nearest neighbor method to calculate the 
background density of dwarf satellites. In the $n$th nearest neighbor
method, the background density $\Sigma_{n}$ of a target galaxy is calculated 
using the following equation,
$$\Sigma_{n}={n \over{\pi r_{p,n}^2}} $$
where $r_{p,n}$ is the projected distance from a target galaxy to the $n$th
nearest neighbor galaxy. There are two constraints for selecting the
neighbor galaxies for the derivation of the background density. One is
the limiting magnitude (${M_{r}}^{\ast}$) of galaxies to be searched for,
and the other is the linking velocity ($\Delta V^{\ast}$) between the target
galaxy and its neighbors.  We used two values of ${M_{r}}^{\ast}$. One
is ${M_{r}}^{\ast}$=-15.2 which is the absolute magnitude of a galaxy with
an apparent magnitude $r=17.77$ at $z=0.01$ and the other 
is ${M_{r}}^{\ast}$=-20.6 which corresponds
to the absolute magnitude of $L^{\ast}$ galaxies for the Schechter luminosity
function fitted to the SDSS galaxies at $z <0.01$ \citep{ann15}.
We adopt $\Delta V^{\ast}$=1000 km s$^{-1}$. 
For $n$, there is no physical ground to select $n$, but, as \citet{mul12}
demonstrated, a small value of $n$ yields a high spatial resolution but with
a low signal-to-noise ratio while a large value of $n$ gives rise to a 
smoothed local background density with a low spatial resolution. We used
$n=5$ for both values of ${M_{r}}^{\ast}$. The background density
derived with ${M_{r}}^{\ast}$=-15.2 gives a higher spatial resolution than
that derived with ${M_{r}}^{\ast}$=-20.6. We use the terminology local
background density ($\Sigma_{L}$) and global background density ($\Sigma_{G}$) 
for the background densities derived with ${M_{r}}^{\ast}$=-15.2 
and ${M_{r}}^{\ast}$=-20.6, respectively. The mean distances to the 5th 
nearest neighbor galaxies are 0.65 Mpc and 9.7 Mpc, respectively 
for $\Sigma_{L}$ and $\Sigma_{G}$.

\begin{table}{t}
\begin{center}
\centering
\caption{Probabilities of the K-S test for 6 sub-types of dwarf galaxies}
\doublerulesep2.0pt
\renewcommand\arraystretch{1.5}
\begin{tabular}{rcccccc}
\hline \hline
\multicolumn{7}{c}{local background density} \\
\hline
Type & dS0 & dE & dE$_{bc}$ & dSph & dE$_{blue}$ & dI \\
\hline
dS0 &   1.000 &   0.617 &   0.436 &   0.615 &   0.031 &   0.026\\
 dE &   0.617 &   1.000 &   0.035 &   0.144 &   0.000 &   0.000\\
 dE$_{bc}$  &   0.436 &   0.035 &   1.000 &   0.283 &   0.027 &   0.007\\
 dSph  &   0.615 &   0.144 &   0.283 &   1.000 &   0.059 &   0.047\\
 dE$_{blue}$  &   0.031 &   0.000 &   0.027 &   0.059 &   1.000 &   0.111\\
 dI  &   0.026 &   0.000 &   0.007 &   0.047 &   0.111 &   1.000\\
\hline
\multicolumn{7}{c}{global background density} \\
\hline
Type & dS0 & dE & dE$_{bc}$ & dSph & dE$_{blue}$ & dI \\
\hline
dS0  &   1.000 &   0.000 &   0.057 &   0.001 &   0.031 &   0.032\\
 dE  &   0.000 &   1.000 &   0.038 &   0.666 &   0.023 &   0.003\\
 dE$_{bc}$  &   0.057 &   0.038 &   1.000 &   0.283 &   0.432 &   0.621\\
 dSph  &   0.001 &   0.666 &   0.283 &   1.000 &   0.171 &   0.037\\
 dE$_{blue}$ &   0.031 &   0.023 &   0.432 &   0.171 &   1.000 &   0.125\\
 dI  &   0.032 &   0.003 &   0.621 &   0.037 &   0.125 &   1.000\\
\hline
\end{tabular}
\end{center}
\end{table}

\begin{figure}
\includegraphics[width=1\columnwidth]{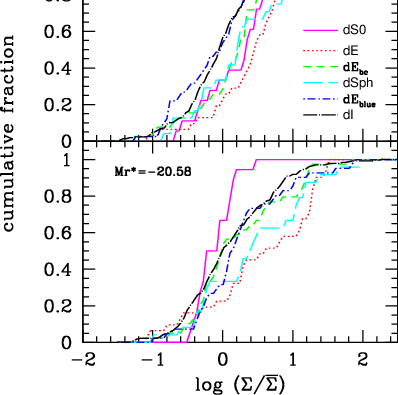}
\caption{Cumulative fraction of dwarf satellites as a function of $\Sigma$.
The background density of local environment ($\Sigma_{L}$) is presented in 
the upper panel and that for the global environment ($\Sigma_{G}$)
is given in the lower panel.
The cumulative fractions of the five sub-types of dwarf elliptical-like
galaxies and dwarf irregular galaxies are plotted by different lines:
solid line (dS0), dotted line (dE), short-dashed line (dE$_{bc}$),
long-dashed line (dSph), dot and short-dashed line (dE$_{blue}$, and
dot and long dashed line (dI). \label{fig7}}
\end{figure}

Figure 7 shows the cumulative fraction of dwarf satellites as a 
function of the background density normalized by the mean background
density ($\Sigma/\bar{\Sigma}$). The local background density ($\Sigma_{L}$)
is plotted in the upper panel and the global background density ($\Sigma_{G}$)
in the lower panel. There are some differences in the cumulative distributions
of $\Sigma_{L}$ between early-type satellites (dS0, dE, dSph, and dE$_{bc}$)
and late-type satellites (dE$_{blue}$ and dI). However, as shown in the 
probability of the K-S test in Table 3, the majority of these differences are 
statistically insignificant because they have probability larger than 
the significance level of $\alpha=0.05$. Among the early-type dwarf satellites, 
dS0, dE, and dE$_{bc}$ galaxies have $\Sigma_{L}$ significantly different from 
those of late-type dwarfs (dE$_{blue}$ and dI). The $\Sigma_{L}$ distribution
of dSph galaxies are significantly different from dI galaxies but the 
significance level is marginal for dE$_{blue}$ galaxies. 
It is of interest to see the $\Sigma_{L}$ distribution of dE$_{bc}$ galaxies.
It is significantly different from dE galaxies as well as dE$_{blue}$ and dI 
galaxies. This means that they are likely to be located in the regions with 
intermediate local background density. 

On the other hand, the distributions of $\Sigma_{G}$ show no such a division
between early-type dwarfs and late-type dwarfs due to different $\Sigma_{G}$ 
distributions of dS0 and dE$_{bc}$ satellites. In particular, the dS0 
satellites have a narrow range of $\Sigma_{G}$, confined to 
$-0.5 < log (\Sigma_{G}/\bar{\Sigma_{G}}) < 0.5$, which makes their 
$\Sigma_{G}$ much different from others. The dE$_{bc}$ satellites have
$\Sigma_{G}$ similar to dE$_{blue}$ and dI. 
Since $\Sigma_{G}$ reflects the galaxy distribution
associated with the large scale structures while $\Sigma_{L}$ reflects the 
galaxy distribution nearby, i.e., galaxies in and around the satellite
system they belong to, it is of interest to
understand why the two background densities, $\Sigma_{L}$ and $\Sigma_{G}$,
of dS0 satellites are so different. The dE$_{bc}$ satellites also show quite
different density distributions for $\Sigma_{L}$ and $\Sigma_{G}$. 
The $\Sigma_{L}$ distribution of dE$_{bc}$ satellites is similar to that
of dSph satellites except for an excess near 
$log (\Sigma_{L}/\bar{\Sigma_{L}})=0.7$ while the $\Sigma_{G}$ of dE$_{bc}$
satellites is similar to that of dE$_{blue}$ and dI of which $\sim60\%$ are
located in the under-dense regions ($\Sigma_{G} < \bar{\Sigma_{G}}$).

\begin{table}{t}
\begin{center}
\centering
\caption{Probabilities of the K-S test for dwarf satellites of late-type hosts}
\doublerulesep2.0pt
\renewcommand\arraystretch{1.5}
\begin{tabular}{rcccccc}
\hline \hline
\multicolumn{7}{c}{local background density} \\
\hline
Type & dS0 & dE & dE$_{bc}$ & dSph & dE$_{blue}$ & dI \\
\hline
dS0 &   1.000 &   0.034 &   0.057 &   0.053 &   0.329 &   0.183\\
 dE &   0.034 &   1.000 &   0.192 &   0.690 &   0.004 &   0.001\\
 dE$_{bc}$  &   0.057 &   0.192 &   1.000 &   0.261 &   0.058 &   0.024\\
 dSph  &   0.053 &   0.690 &   0.261 &   1.000 &   0.016 &   0.011\\
 dE$_{blue}$  &   0.329 &   0.004 &   0.058 &   0.016 &   1.000 &   0.274\\
 dI  &   0.183 &   0.001 &   0.024 &   0.011 &   0.274 &   1.000\\
\hline
\multicolumn{7}{c}{global background density} \\
\hline
dS0  &   1.000 &   0.056 &   0.460 &   0.095 &   0.393 &   0.576\\
 dE  &   0.056 &   1.000 &   0.024 &   0.375 &   0.000 &   0.000\\
 dE$_{bc}$  &   0.460 &   0.024 &   1.000 &   0.300 &   0.650 &   0.498\\
 dSph  &   0.095 &   0.375 &   0.300 &   1.000 &   0.046 &   0.011\\
 dE$_{blue}$ &   0.393 &   0.000 &   0.650 &   0.046 &   1.000 &   0.089\\
 dI  &   0.576 &   0.000 &   0.498 &   0.011 &   0.089 &   1.000\\
\hline
\end{tabular}
\end{center}
\end{table}

Figure 8 shows the cumulative distributions of the dwarf satellites divided
into the host galaxy morphology as a function of the background densities, 
$\Sigma_{L}$ in the left panels and $\Sigma_{G}$ in the right panels.
There is a large difference between the $\Sigma_{L}$ and $\Sigma_{G}$.
The mean $\Sigma_{L}$ of the satellites of early-type galaxies is similar to
that of late-type galaxies, whereas the mean $\Sigma_{G}$ of satellites of 
late-type galaxies is much higher than that of the early-type galaxies 
if we ignore the dS0 satellites of late-type galaxies. On average,
the fraction of satellites of early-type galaxies located in the under-dense
regions ($\Sigma_{G} < \bar{\Sigma_{G}}$) is $\sim0.6$ while the fraction
of satellites of late-type galaxies located
in the under-dense region ($\Sigma_{G} < \bar{\Sigma_{G}}$) is $\sim0.3$.
In the satellite systems hosted by early-type galaxies, the distributions
of satellite background densities show no significant difference among
the sub-types, whereas in the satellite systems hosted by late-type galaxies,
there are significant differences between  
early-type satellites and late-type satellites if we exclude the dS0 
galaxies which have a narrow range of the local background densities around
$log (\Sigma_{L}/\bar{\Sigma_{L}}) \approx -0.5$. 
This feature is more pronounced in the distributions of the global 
background densities ($\Sigma_{G}$) where we see that the $\Sigma_{G}$
of dE and dSph satellites of late-type galaxies are significantly different
from those of dE$_{blue}$ and dI satellites. The dE$_{bc}$ satellites show
$\Sigma_{G}$ distribution similar to late-type satellites. 
We summarize the results of the K-S test to show the significance of the
differences between the sub-types of dwarf satellite galaxies in Table 4.
We omit the results for the satellites of 
early-type hosts because there is no significant difference at all.
The $\Sigma_{G}$ distribution of dE$_{bc}$  satellites
may be related to the origin of the cold gas
in the central regions of dE$_{bc}$ galaxies. That is, if the cold gas of
dE$_{bc}$ satellites is accreted from the intergalactic medium, dE$_{bc}$ 
galaxies are likely to be found in the regions where cold gas are abundant.
We suppose that cold gas is more abundant in the under-dense regions than 
in the over-dense regions.

\begin{figure}
\includegraphics[width=1\columnwidth]{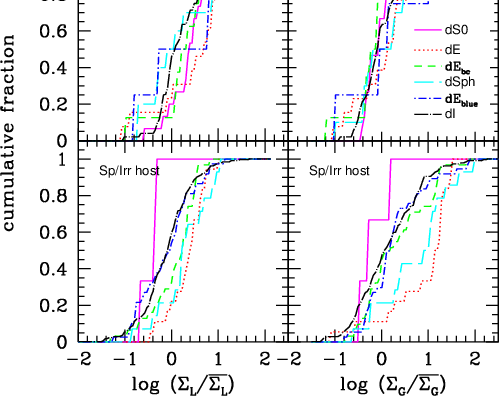}
\caption{Cumulative fraction of dwarf satellites as a function of
$log (\Sigma/\bar{\Sigma})$. $\Sigma_{L}$  is given in the left panels and
$\Sigma_{G}$ in the right panels. Satellites in early-type hosts
are presented in the upper panels while those
for late-type hosts are displayed in the lower panels.
The cumulative fractions of the five sub-types of dwarf elliptical-like
galaxies and dwarf irregular galaxies are plotted by different lines:
solid line (dS0), dotted line (dE), short-dashed line (dE$_{bc}$),
long-dashed line (dSph), dot and short-dashed line (dE$_{blue}$), and
dot and long dashed line (dI). \label{fig8}}
\end{figure}

The dependence of the background densities on the host and satellite
morphology, demonstrated as the differences in the distributions of the
background densities shown in Figure 8, implies that the main mechanism
to determine the satellite morphology depends on the host morphology as well 
as the background densities of satellites. The morphology of a satellite
of an early-type host galaxy seems to be mostly determined by the 
interactions with the host galaxy while the morphology of a satellite of
a late-type host is more affected by the environment 
where they formed. The
difference between $\Sigma_{L}$ and $\Sigma_{G}$ of satellites hosted by 
early-type galaxies seem to be caused by the relative importance of local and 
global background densities compared with the effect of interactions between
host and satellites.
The hydrodynamical and tidal interactions between host and satellites
seem to be the main cause that determines the morphology of a satellite 
around an early-type host but the local environment, represented by 
$\Sigma_{L}$, may affect the satellite morphology somewhat. On the contrary,
the global environment, represented by $\Sigma_{G}$, does not affect the
satellite morphology in the systems hosted by early-type galaxies.

On the other hand, the satellite morphology seems to be significantly affected
by the local and global background densities in the satellite systems hosted by 
late-type galaxies. If we exclude the dS0 satellites of late-type 
hosts (three galaxies only),
there is a clear difference in the distribution of the background
densities between early-type satellites and late-type satellites. 
In particular, the distribution of $\Sigma_{G}$ of dE and dSph satellites
are significantly different from those of dE$_{blue}$ and dI satellites.
It suggests that most dE and Sph satellites of late-type galaxies,
which are located at large galactocentric distances, may have
their current morphology before
they fall into the satellite system. They may be primordial ones formed in the
dense regions in the early history of galaxy formation or they got the current
morphology by the frequent interactions with neighbor galaxies, expected in the 
dense regions.

The argument that early-type satellites, dE and dSph in particular, 
of early-type galaxies are transformed from late-type 
galaxies (spirals and irregulars) after they fall into the satellite systems
and dE and dSph satellites of late-type galaxies have the current morphology 
before they come into the satellite systems is supported by their spatial
distributions shown in Figure 6. More than $70\%$ of dE and dSph satellites
of early-type galaxies are located at $r_{p} < 0.15$ Mpc where 
strong hydrodynamical and tidal interactions between the host and
satellites are expected. It is highly contrasted with the spatial distributions
of dE and dSph satellites of late-type galaxies which show that $\sim30\%$ 
of dE and dSph galaxies are located at $r_{p} > 0.25$ Mpc where we expect
little interaction between host and satellites.

\section{Summary and Discussion}

We have studied the dependence of satellite morphology on the environment
represented by the projected distance from host galaxy ($r_{p}$) and the
background densities ($\Sigma$) derived by the $n$th nearest neighbor method
using the visually classified morphological types of the nearby galaxies
at $z \lesssim 0.01$ \citep{ann15}. We have derived a sample of isolated 
satellite systems using variable linking distance ($LD$) and fixed linking
velocity of $\Delta V^{\ast}$=500 km s$^{-1}$. The $LD$ we used is the sum of
the virial radii of host and satellite. The use of variable $LD$ and a fairly
small value of $\Delta V^{\ast}$ allows us to minimize the inclusion of 
interlopers in our satellite sample.
The number of isolated satellite systems and the number of satellites
therein depend on the constraints for the isolation of host galaxies.
We consider a galaxy is an isolated host galaxy if there is no neighbor
galaxy that has a comparable luminosity, i.e., no neighbor galaxy with 
$M_{r} < M_{r,host} + \delta M$ within $r_{p}=r_{vir}+r_{virnei}$. The number 
of isolated satellite systems using $\delta M=1$ is 346 and 
the mean number of satellites in an isolated satellite system is 2.4.
We have explored other values of $\delta M$ such as 0.5, 1.2, and 1.5 
to see the effect of $\delta M$ on the correlation between the satellite 
morphology and its environment and found that $\delta M$ between 0.5 and 1.5 
does not affect the results significantly. 
 
We have found that the detailed morphologies of dwarf satellite galaxies 
depend strongly on the host morphology and $r_{p}$. The local background 
density, derived from galaxies brighter than $M_{r}=-15.24$ which is
the absolute magnitude of a galaxy with $r=17.77$ at $z=0.01$, seems
to play insignificant role on the morphology of dwarf satellite galaxies.
However, the global background density, derived from galaxies brighter than
$M_{r}=-20.58$, which corresponds to the absolute magnitude of the $L^{\ast}$
galaxy in the luminosity function of the local galaxies ($z<0.01$) seems to
affect the morphology of dwarf satellites of late-type galaxies. The global
background density of dE$_{bc}$ satellites similar to dE$_{blue}$ and dI
satellites rather than dE and dSph satellites. Since dE$_{bc}$ and dE galaxies
have similar luminosities which are on average $\sim1.5$ mag brighter than 
those of dE$_{blue}$ and dI galaxies, the environment, represented by the
background densities, plays more significant role on the morphology of dwarf
satellites than the mass of dwarf satellites. 

The spatial distribution of dwarf satellites suggests that the environmental 
quenching driven by the ram pressure stripping \citep{gun72} of the cold gas
in satellite galaxies is operating in the dwarf satellites, especially
for the dwarf satellites of the early-type galaxies. The tidal stirring of 
stars of satellites \citep{may01} may change the structure of galaxies
from disk shapes to ellipsoidal/spheroidal shapes. This is the reason why dE 
and dSph galaxies are preponderant in the vicinity of early-type host galaxies.
The local and global background densities of dwarf satellites of early-type 
galaxies are irrelevant to the satellite morphology. On the other hand,
a significant fraction of dE and dSph satellites of late-type galaxies seems
to be primordial or pre-processed objects because they are located in the
outer parts of the satellite systems where environmental quenching is supposed
to be ineffective. 

Among the satellite spatial distributions, the spatial
distribution of dE$_{bc}$ satellites of early-type galaxies is of special
interest because it may suggest the origin of the cold gas and the dominant
mechanisms for the environmental quenching in the isolated satellite systems. 
The dE$_{bc}$ galaxies are characterized by the blue core which indicates the
presence of the young stellar populations that require cold gas.
There are two explanations for the origin of the cold gas in the 
dE$_{bc}$ satellites if we exclude the recycled gas due to stellar evolution
which seems to be less likely \citep{hal12}.
One is the leftover material after transformation from
late-type galaxies to dE-looking galaxies and the other is the intergalactic 
material accreted after they formed as dE galaxies  \citep{hal12}.
Accreted cold gas can fall into the center of a dwarf galaxy without
being evaporated by the hot corona gas \citep{nipo07}.
The most pronounced feature of their spatial distribution is the preponderance
of dE$_{bc}$ satellites in the outer parts of the satellite systems,
more than $\sim60\%$ of dE$_{bc}$ satellites at $r_{p} >0.3$ Mpc. This 
preference of outlying locations for the dE$_{bc}$ satellites of early-type
galaxies is consistent with the locations of the dE$_{bc}$ galaxies in the
Virgo cluster \citep{hal12}.
If they are leftover material which survives from the ram pressure of the
hot corona of early-type hosts owing to their large separation from
the host galaxies, we do not expect any dependence of dE$_{bc}$
satellites on the background density. On the other hand, if the cold gas are 
intergalactic material accreted before or after becoming the satellites 
of early-type galaxies, we expect some dependence of dE$_{bc}$ satellites on
the background density. In this regard, the global density distribution of 
dwarf satellites of early-type galaxies, shown in the upper-right panel of 
Figure 8, seems
to be informative. The fraction of galaxies that have a global background 
density smaller than the mean value is largest for dE$_{bc}$ 
satellites ($\sim85\%$. This fraction of dE$_{bc}$ satellites is about two
times larger than that of dE satellites. Thus, it seems more plausible to
assume that the cold gas in dE$_{bc}$ satellites is the accreted
intergalactic material. We further suppose that most of dE$_{bc}$ satellites
become satellites of early-type galaxies recently because the amount of cold
gas is not large enough to last for a long time. The spatial distribution of 
dE$_{bc}$ satellites, characterized by the preponderance of dE$_{bc}$ 
satellites in the outer parts of the satellite system, implies that
the ram pressure stripping \citep{gun72} is operating in the vicinity of 
early-type hosts. We think that cold gas in dE-looking satellites is easily
removed by the ram pressure of early-type host galaxies when they are close
enough. It is the reason for the lack of dE$_{bc}$ satellites in the 
vicinity of early-type hosts. 

An example of the results of the hydrodynamical interaction between host
and satellites is the morphology conformity between host and satellites
shown in Figure 5. The hydrodynamical interaction between the hot corona 
gas of a host galaxy and the cold gas of satellite, especially in the satellite
systems hosted by early-type galaxies, removes the cold gas from the 
satellites to transform star forming late-type galaxies into quiescent
early-type ones. Tidal interactions play a critical role on transforming
the disk-like structure into ellipsoidal/spheroidal shapes.
The lack of dE$_{blue}$ galaxies at $r_{p} <0.1$ Mpc from early-type
hosts also demonstrates the importance of the hydrodynamical
interactions in the satellite systems when a host galaxy has a large hot 
corona which is expected to be present in luminous early-type galaxies.
The hydrodynamical interaction between a late-type host galaxy and its 
satellites, which leads to ram pressure stripping of cold gas in the satellite
galaxies, seems to be less effective because the hot corona of the late-type
galaxy is less luminous than that of the early-type galaxy.


\acknowledgments

This work was supported partially by the NRF Research grant 2015R1D1A1A09057394.

\clearpage{}
\end{document}